# Surfactant-Free Polar-to-Non-Polar-Phase Transfer of Exfoliated $MoS_2$ Two-Dimensional Colloids

Emerson Giovanelli,*[a] Andres Castellanos-Gomez[a] and Emilio M. Pérez*[a]

[a] IMDEA Nanociencia, Calle Faraday 9, Ciudad Universitaria de Cantoblanco, 28049, Madrid, Spain
* E-mail: emerson.giovanelli@imdea.org; emilio.perez@imdea.org

**Abstract:** *Exfoliation of lamellar materials into their corresponding layers represented a breakthrough, due to the outstanding properties arising from the nanometric thickness confinement. Among the cleavage techniques, liquid-phase exfoliation is now on the rise because it is scalable and leads to easy-to-manipulate colloids. However, all appropriate exfoliating solvents exhibit strong polarity, which restrains a lot the scope of feasible functionalization or processing of the resulting flakes. Here we propose to extend this scope, demonstrating that nanosheets exfoliated in a polar medium can be properly dispersed in a non-polar solvent. To that purpose, we prepared suspensions of molybdenum disulfide flakes in isopropanol/water and developed a phase transfer of the nanosheets to chloroform via precipitation and redispersion/centrifugation sequences, without any assisting surfactant. The colloidal stability of the nanosheets in chloroform was found to be governed by their lateral dimensions and, although lower than in polar media, proved to be high enough to open the way to subsequent functionalization or processing of the flakes in non-polar medium.*

**Introduction**

The exfoliation of bulk layered materials into two-dimensional (2D) nanosheets has triggered a significant interest of the scientific community since 2004 and the isolation of one-atom-thick graphene by mechanical exfoliation of graphite, using the so-called 'adhesive-tape method'.[1] Representative and relevant examples of 2D materials beyond graphene prepared by the adhesive-tape method are transition metal dichalcogenides (TMDCs), and most particularly molybdenum disulfide ($MoS_2$). The derived nanosheets, down to the single layer, actually turned out to be valuable not only to act as a miniaturized form of the bulk semiconductor species but especially for the exciting optoelectronic properties (*e. g.* the indirect-to-direct bandgap transition leading to fluorescence emission) coming from the thickness confinement.[2-3]

Among the exfoliation techniques available to date, the adhesive-tape mechanical cleavage certainly remains the method of choice for fundamental studies and prototyping. Nevertheless, the need for reliable mass production procedures has strongly pushed the development of wet-phase-type processing.[4-6] Such a technique is based on the ultrasound irradiation of a layered material as a micrometric powder dispersed in a determined solvent, and results in easy-to-handle colloidal suspensions.[7] Over the last few years, this approach, known as liquid-phase exfoliation (LPE), has largely extended its scope in terms of both the process efficiency[5, 8-11] and the materials' variety.[12-17]

One of the next challenges in the field of layered materials wet processing consists in further functionalizing the colloidal nanosheets to modify their physico-chemical properties (charge state, solvent dispersibility, n/p doping, anchoring functions to create heterostructures…), taking advantage of the liquid-phase dispersion. The functionalization can be achieved via covalent bonds or supramolecular assembly (ligand or metal coordination, physisorption) and some substantial work has already been performed in that regard, but almost exclusively limited to polar phases.[18-22] Polar solvents are actually strongly preferred as exfoliation media: from the cohesive forces existing within these solvents at the microscopic scale (dipole-dipole attraction, hydrogen bonding, van der Waals forces) derive high surface tensions that are suitable for the layer separation, while the heteroatoms they bear (mainly N and O) confer them coordination sites that favor the colloidal dispersion of the nanosheets.[23] Additionally, in the case of $MoS_2$ for example, LPE proved to lead to charged flakes,[12, 24-26] which are consequently better dispersed in a polar medium, due to solvation effects and better charge separation that enhances stabilizing electrostatic repulsion. In a more general approach, liquid-phase exfoliation can thus be described in the framework of the generalized solubility theory extended to 2D materials.[27]

Hence, polar solvents play a decisive role in breaking the van der Waals interaction existing in-between two-dimensional layers, and definitely contribute to the colloidal stability of the nanosheets. That is the reason why the phase transfers investigated so far only consisted in exchanges between polar solvents.[28] However, such polar environments prevent the chemist from exploring a wide range of functionalization reactions with a variety of hydrophobic molecular reagents or species, due to the lack of solubility of the latter. Furthermore, small polar molecules used as solvents generally possess higher boiling points than the non-polar ones, owing to the above mentioned cohesive forces, which can also limit the colloid processing into devices (slow evaporation, remaining solvent residues, bad wetting of the substrate...).

Here we propose a simple way to progressively transfer the $MoS_2$ nanosheets initially prepared in a polar solvent mixture of isopropanol (*i*PrOH, $\varepsilon_r$ = 18.3 at 20 °C, µ = 1.66 D, b.p. 82 °C) and water ($H_2O$, $\varepsilon_r$ = 79.7 at 20 °C, µ = 1.87 D, b.p 100 °C) to chloroform ($CHCl_3$), a non-polar solvent due to its dielectric constant lower than 5 ($\varepsilon_r$ = 4.8 at 20 °C, µ = 1.1 D), that is able to solubilize most of the species that are not soluble in polar solvents and exhibits a low boiling point (b.p. 61 °C).[29] Through centrifugation/redispersion sequences, we thus managed to get nanosheet colloidal suspensions in chloroform from a variety of exfoliated $MoS_2$ samples, and without resorting to surfactants to stabilize the flakes. All the colloids were characterized by UV-Visible and fluorescence spectroscopies and the colloidal stability of the obtained suspensions was investigated in detail.

## Results and Discussion

## Probe-type exfoliation of MoS$_2$ in polar *i*PrOH/H$_2$O mixture

All of the MoS$_2$ LPE experiments presented here have been carried out with the same concentration in MoS$_2$ powder in the initial dispersion (1 mg.mL$^{-1}$) and the same solvent mixture, namely *i*PrOH/H$_2$O 7/3 (v/v). Note that this mixture of polar solvents has been found to best match both the surface tension polar and dispersive components of $MoS_2$, thus making the exfoliation process and the stabilization of the resulting nanosheets more effective in that medium than in any other commonly used polar solvent.[10] The preceding conditions have been chosen according to a previously published work of our group on the optimization of $MoS_2$ exfoliation to fluorescent monolayers via bath sonication.[30] In the present case, the exfoliation was performed with the help of an ultrasonic probe, varying the sonication time only, from 1 min to 60 min. Each experiment was performed twice to check reproducibility. Note that such experiments produce nanosheets that are usually a few hundreds of nanometres in size;[30] as a consequence, scattering is no longer negligible and the UV-Visible characterization of the corresponding colloids gives extinction spectra that sum both the light absorption and scattering due to the sample. The extinction spectra of the nanosheet colloidal suspensions obtained after centrifugation are presented in Figure 1a-b.

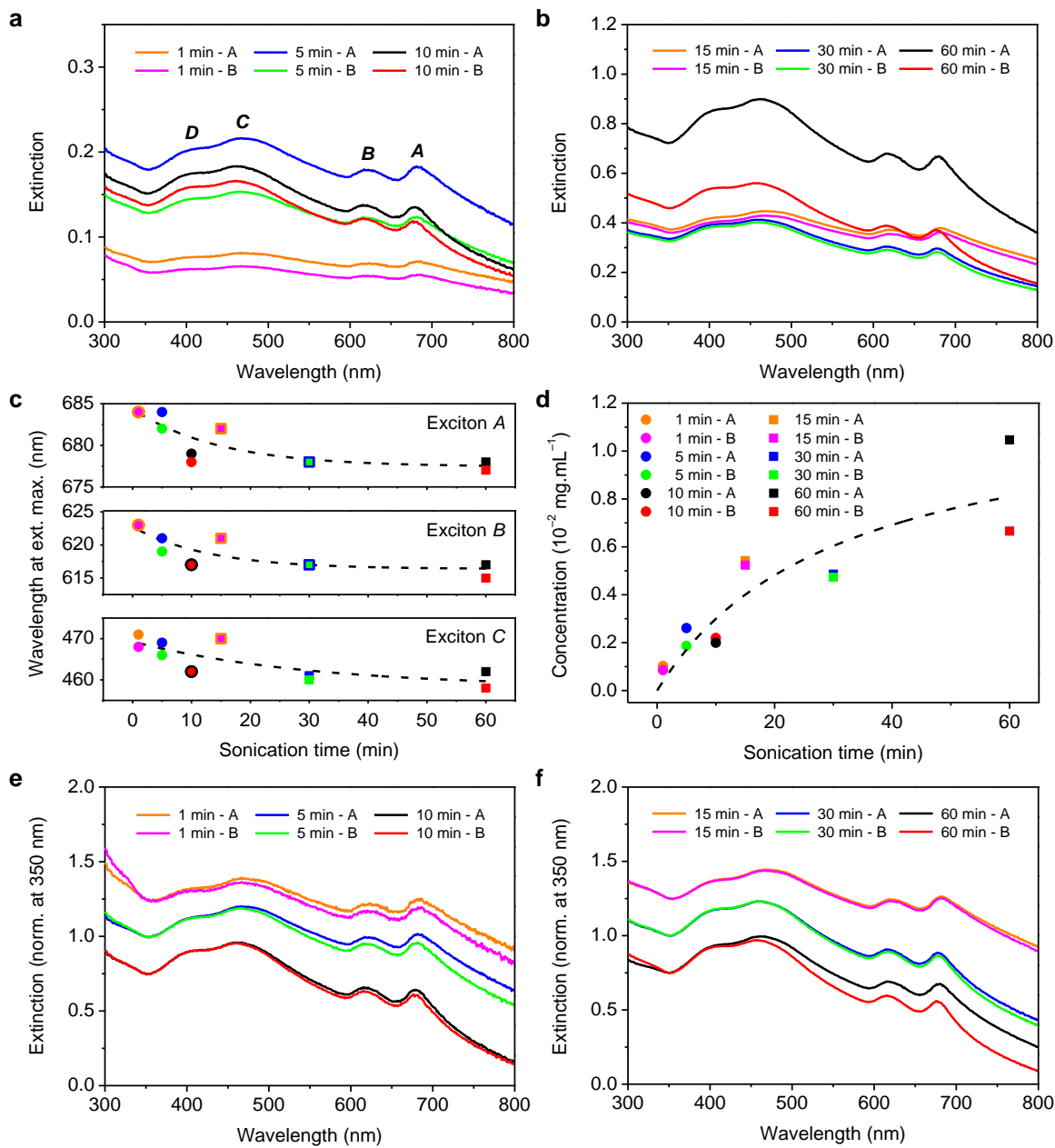

**Figure 1. UV-Visible spectroscopy of MoS$_2$ colloids obtained after exfoliation of MoS$_2$ powder dispersions in *i*PrOH/H$_2$O 7/3 (v/v) by probe sonication.** Each sample is referred to as the corresponding sonication time, the letter A or B indicating the first and second replicates respectively. **a.** Extinction spectra of as-obtained suspensions in *i*PrOH/H$_2$O 7/3 after the following sonication times: 1 min (orange and pink lines), 5 min (blue and green lines) and 10 min (black and red lines). The different absorption bands are attributed to the corresponding excitons *A*, *B*, *C* and *D*. **b.** Extinction spectra of as-obtained suspensions in *i*PrOH/H$_2$O 7/3 after the following sonication times: 15 min (orange and pink lines), 30 min (blue and green lines) and 60 min (black and red lines). **c.** Wavelengths at extinction maxima for *A*, *B* and *C* excitons as a function of the sonication time. Each data point corresponds to one experiment identified in **d**. Trend curves are represented as dashed black lines. **d.** Extinction value at 350 nm as a function of the sonication time for all experiments and corresponding trend (black dashed line). **e.** Spectra shown in **a** normalized at 350 nm (pairs of replicates separated for clarity). **f.** Spectra shown in **b** normalized at 350 nm (pairs of replicates separated for clarity).

All of the spectra show the typical features of $MoS_2$ nanosheet suspensions, namely four absorption bands corresponding respectively to *A* and *B* excitons, and *C* and *D* excitons in the order of decreasing wavelengths.[31] *A* and *B* excitons originate both from the direct bandgap transition at the K point of the Brillouin zone, the degeneracy of the top of the valence band being partially lifted due to spin-orbit coupling.[2, 32] As for *C* and *D* excitons, they correspond to higher energy transitions favored by the band edges being parallel to each other between the Γ and the Λ points of the Brillouin zone (so-called 'band nesting effect').[33-35]

As previously observed in other LPE experiments,[30] the positions of the respective band maxima slightly shift toward the blue as the sonication time increases: from 684 to 677 nm for exciton *A*, from 623 to 615 nm for exciton *B*, and from 471 to 458 nm for exciton *C* (Figure 1c). Only the *D* exciton shoulder does not show any noticeable shift as a function of the sonication time and remains centered at ~405 nm. Fractional sedimentation experiments carried out by Coleman and co-workers demonstrated that such blue-shifts correlate with an increase in the weight of the flakes, *i.e.* in their lateral dimensions and/or thickness.[14] They showed that within a typical $MoS_2$ nanosheet colloid extinction spectrum, the scattering contribution exhibits red-shifted bands as compared to the pure absorption contribution. As samples containing small and thin nanosheets exhibit far less scattering than those made of thick and/or large flakes, absorption dominates over scattering in their extinction spectra and the respective band maxima appear at shorter wavelengths. Additionally, the proper absorption maxima prove to blueshift as the thickness of the flakes decreases, due to quantum confinement or size effect,[2, 14, 33, 36-38] which can further accentuate the same trend. As an additional proof, we observed by transmission electron microscopy (TEM) the nanosheets obtained after 1 min (100-200 nm in size, Figure 2a,b and S1) and 60 min sonication (≥ 500 nm, Figure 2c,d and S2). We thus confirm that in our experimental conditions, the longer the sonication time is, the smaller and the thinner the colloidal flakes are.

Previous studies revealed that the extinction coefficient at the local minimum in the UV part of the extinction spectrum, observed in our case at 350 nm, is rather size-independent, the absorption dominating over scattering.[14] So we could use the determined mass extinction coefficient (6,900 $L.g^{-1}.m^{-1}$) to estimate the concentration in suspended material after each exfoliation experiment (Figure 1d). The results show colloid concentrations in the 1-10 $mg.L^{-1}$ range, with an increase in suspended $MoS_2$ with the sonication time, following an exponential decay trend. Further comparison with previously published experiments[30] confirms the ultrasonic probe exfoliation is more efficient than its bath counterpart: at similar irradiation times, the former produces twice to three times as much material as the latter.

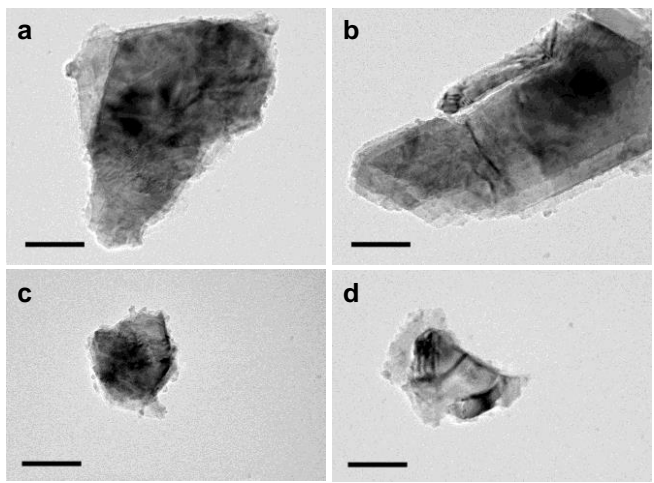

**Figure 2. Representative TEM micrographs of the nanosheets obtained after exfoliation in *i*PrOH/H$_2$O 7/3 and centrifugation. a, b**: sample '1 min – A'. **c, d**: sample '60 min – A'. Scale bars are 100 nm.

Another part of the spectrum of particular interest is the extinction measured at wavelengths higher than 700 nm. In that range, MoS$_2$ nanosheets do not absorb and the extinction consists of pure scattering;[14] lower measured intensities, which would also correspond to more pronounced spectrum curvatures beyond 700 nm, are thus indicative of smaller-sized flakes. The preceding extinction spectra were therefore normalized at 350 nm (Figure 1e-f and Figure S3) and shown at the same scale to allow their fair comparison. The following trend is observed: in the series of increasing sonication times, the extinction at wavelengths higher than 700 nm – the scattering – tends to decrease. This suggests a decrease in the average nanosheet size, which further confirms the previous conclusion drawn from the blue-shifts of the extinction bands and the TEM observations.

We satisfactorily check the reproducibility of the experiments by comparison of the replicates for each sonication time. Nevertheless, a few observations should be addressed. The original spectra point out noticeable differences in the extinction for the '1 min', '5 min' and '60 min' experiments (Figure 1a-b). The corresponding two replicates show indeed mean relative differences in their extinction values of 20%, 31% and 43% respectively, whereas in the case of the other experiments, the mean relative differences do not exceed 11%. As normalization eliminates the effects due to concentration, Figure 1e-f highlights more specifically the differences in spectral features between the replicates: wavelengths of local extinction maxima and scattering contribution differ slightly within the '1 min' and within the '5 min' experiments, and more remarkably within the '60 min' replicates, along with obvious changes in the band relative intensities. The other three experiments produce samples with identical features. Hence, both representations coincide and reveal differences in both the concentration and the nanosheet composition between the replicates of the '1 min', '5 min' and '60 min' experiments (shortest and longest sonication times). The deviations observed at short sonication times are rather small from an absolute point of view, so they must come from slight variations in the regulation of the centrifugation temperature (modification of the nanosheet sedimentation process) or in the collection of the supernatant after centrifugation (more significant effects at low concentrations). As for the '60 min' experiments, the differences observed are most probably due to significant temperature variations during the exfoliation step: at such sonication times, since the control of the temperature is only external, overheating of one of the samples is likely to have occurred.

Finally, we demonstrated the 2H structure and the absence of degradation of MoS$_2$ nanosheets with the sonication time by Raman spectroscopy. Samples exfoliated for 1 min and 60 min exhibit identical spectra

(Figure S4a-c), showing the two characteristic bands at 378 and 405 cm$^{-1}$ corresponding to in-plane $E^1_{2g}$ and out-of-plane $A_{1g}$ vibration modes of 2H MoS$_2$.[39]

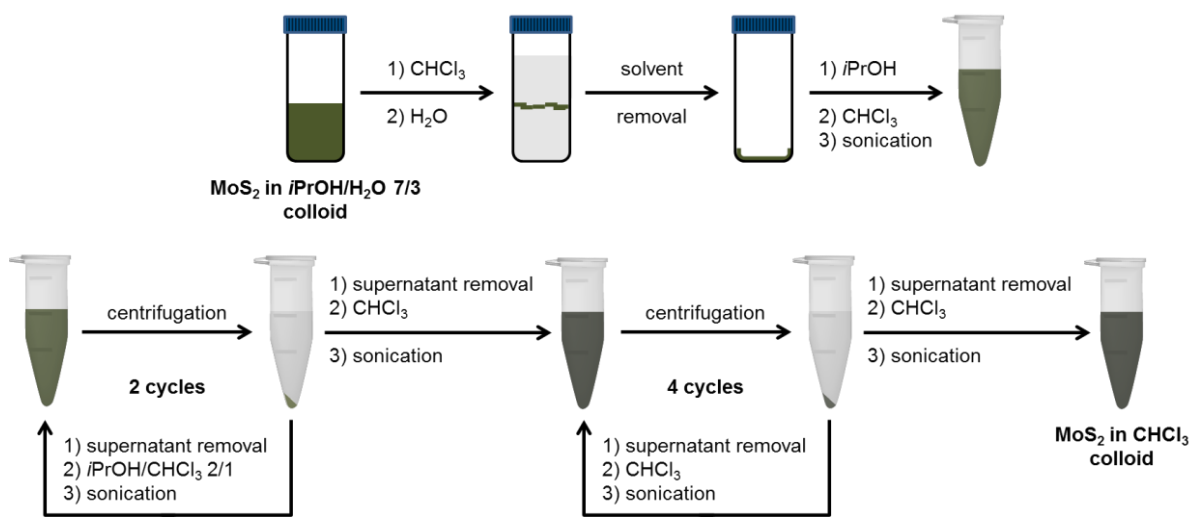

**Figure 3.** Principle of colloidal MoS$_2$ nanosheet phase transfer from *i*PrOH/H$_2$O (7/3, v/v) to chloroform (not to scale).

**Phase transfer of MoS$_2$ nanosheets to chloroform**

MoS$_2$ nanosheets obtained as stable colloids in *i*PrOH/H$_2$O 7/3 are transferred to chloroform via a multi-step process whose principle is illustrated by Figure 3 (see the 'Experimental Section' for all technical details). The procedure consists in progressively removing the polar solvents in favor of the non-polar one and relies on two main stages: the elimination of water and the dispersion of the nanosheets into chloroform. Preliminary experiments showed that thorough removal of water is crucial to get colloidal stability in chloroform, most probably due to the solvents' immiscibility.

First, chloroform is added to the MoS$_2$ suspension in *i*PrOH/H$_2$O until the ensemble demixes, which requires *ca.* half of the initial suspension volume in chloroform. An equal volume of pure water, immiscible with chloroform, is further added to ensure complete phase separation. Such a demixing causes the nanosheets to flocculate at the interface between the two liquid phases which are subsequently removed. The obtained MoS$_2$ residue is redispersed in the minimal volume of *i*PrOH, to which a half part of chloroform is added. As *i*PrOH is miscible with CHCl$_3$, a homogeneous phase is formed. *i*PrOH being its major component, the mixture is able to both solubilize the last traces of water within MoS$_2$ residue and disperse the nanosheets. The role of CHCl$_3$ is to limit the colloidal stability (turbid suspension) so that the water can be washed from the nanosheets through two redispersion/centrifugation sequences in *i*PrOH/CHCl$_3$ 2/1. Then, MoS$_2$ nanosheet sediment is redispersed in CHCl$_3$, and taking advantage of *i*PrOH miscibility with CHCl$_3$, remaining *i*PrOH is washed from the nanosheets via similar redispersion/ centrifugation sequences, this time in pure CHCl$_3$. Due to the lesser abilities of CHCl$_3$ to disperse charges (non-polar solvent, low $\varepsilon_r$), the stability of the colloids is not as high as in polar solvents; as a consequence, high speed centrifugation (18,626 *g*) is enough to correctly separate the nanosheets from the solvent. The washing sequence is performed four times to ensure complete removal of *i*PrOH. It results in MoS$_2$ nanosheet suspensions that are colloidally stable in chloroform. Note that chloroform, which possesses a dipole moment in spite of its very low dielectric constant, may participate in the flake stabilization through ion-dipole interactions with the residual charges remaining onto the nanosheets. Surface chemistry experiments[40] and numerical simulations[41] have also highlighted the existence of stabilizing weak dispersive interactions between MoS$_2$ basal plane and non-polar species.

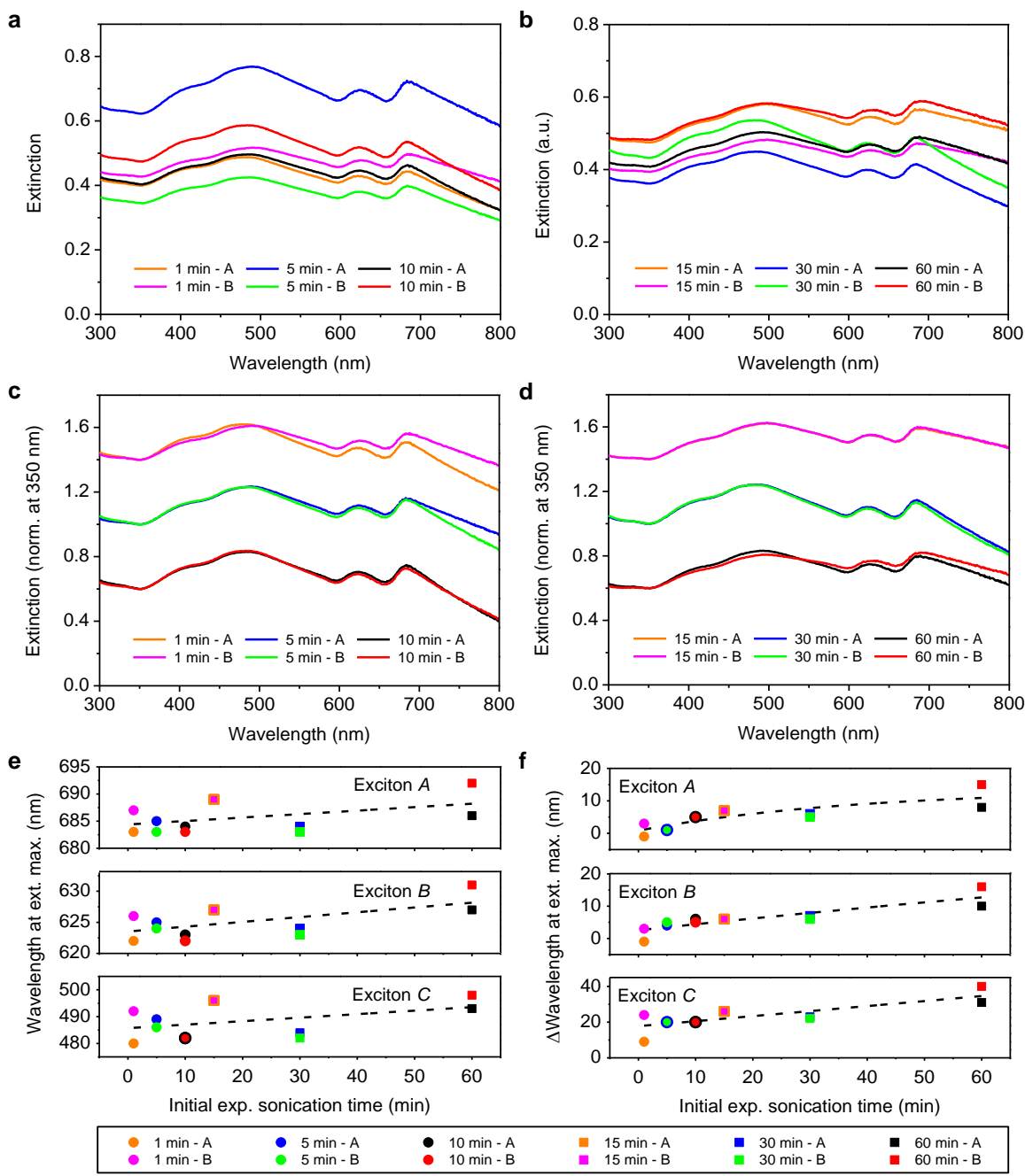

**Figure 4. UV-Visible spectroscopy of exfoliated MoS$_2$ nanosheets after phase transfer into chloroform. a.** and **b.** Extinction spectra of MoS$_2$ nanosheets in colloidal suspension in chloroform after the phase transfer. **c.** Spectra shown in **a** normalized at 350 nm (pairs of replicates separated for clarity). **d.** Spectra shown in **b** normalized at 350 nm (pairs of replicates separated for clarity). As in Figure 1, each sample is referred to as the sonication time used for the initial exfoliation experiment in *i*PrOH/H$_2$O 7/3, associated with the letter A or B, indicating the first and second replicates respectively. **e.** Wavelengths at extinction maxima for *A*, *B* and *C* excitons as a function of the sample, identified by the initial experiment sonication time. **f.** Variation of the wavelengths at extinction maxima in CHCl$_3$ with respect to those in *i*PrOH/H$_2$O for *A*, *B* and *C* excitons as a function of the sample, identified by the initial experiment sonication time. The legend of **e** and **f** is located below the corresponding graphs.

All samples, including replicates, were subjected to the above described phase transfer and characterized by UV-Visible spectroscopy. As the transfer process was also used to concentrate the samples, their respective extinction values could be adjusted to the same range for better comparison. The results are shown in Figure 4a-b. A first overview does not let appear as many variations among the extinction spectra in chloroform as between those of the original polar suspensions (Figure 1a-b). Nevertheless, a further comparison of the normalized spectra (Figure 4c-d and Figure S5) stresses that the scattering contribution, directly observable at wavelengths higher than 700 nm, globally tends to increase as the size of the nanosheets composing the initial polar colloid decreases, from the '1 min' to the '60 min' samples. We can also note that the phase transfer proves to be highly reproducible: the pairs of replicates that exhibited identical normalized spectra in $i$PrOH/H$_2$O ('10 min', '15 min' and '30 min') show spectra in CHCl$_3$ that can be superimposed again; similarly, slight differences persist between the replicates of the other experiments, in spite of analogous spectra.

In order to characterize more precisely the chloroform suspensions, we reported the wavelengths of the local extinction maxima (corresponding to excitons *A*, *B*, and *C*; the *D* exciton remains at 405 nm, as in the polar phase) as a function of the samples, referred to as the sonication time used for their preparation (Figure 4e). The graphs show the trend is opposite to that observed previously in the polar phase: *A*, *B* and *C* excitons in the CHCl$_3$ suspensions slightly redshift as the sonication time increases. The explanation is to be found in how the exciton band wavelengths shift from the polar phase to the non-polar phase suspensions, according to the sonication time initially used to prepare each sample. As depicted in Figure 4f, the phase transfer results in a red-shift of all excitons with respect to the polar phase suspensions. Additionally, this red-shift especially increases for the samples resulting from the longest sonication times and corresponding to the smallest nanosheets in the polar suspensions (see preceding section).

Given these features of the chloroform suspensions extinction spectra, we ruled out any significant solvatochromic effect on the absorption bands. Indeed, we expect it would have affected all samples to a similar extent. Furthermore, the absence of shift in the photoluminescence wavelength (see the final subsection about fluorescence) does not point to such an effect of chloroform. Raman analysis of the '1 min' and '60 min' samples does not evidence any structural change of the nanosheets either (Figure S4a, b, d-f). As a consequence, we would rather attribute the red-shift of the excitonic bands to an increase in the scattering contribution. Note that the proper scattering bands redshift and broaden as the refractive index increases[42] ($n_{i\text{PrOH/H2O 7/3}}$ = 1.374;[43] $n_{\text{CHCl3}}$ = 1.444).[29] Nevertheless, this does not explain the observed differential red-shift (along with the differential increase of the scattering contribution) as the nanosheet size decreases (Figure 4f). We therefore hypothesize that in chloroform, the smallest nanosheets aggregate more easily than the largest ones, thus increasing the scattering cross-section in a more substantial way. This assumption is consistent with our TEM observations of the '1 min' (Figure S6) and '60 min' (Figure S7) samples after the phase transfer. Such a behavior can be rationalized using the Derjaguin-Landau-Verwey-Overbeek (DLVO) theory on colloidal stability[44] adapted to nanoplate-like structures.[45-46] The theory states that in the absence of gravity effects (which is the case here owing to the centrifugation removal of poorly exfoliated material) and steric repulsion (no ligands on the nanosheets), the stability of charged colloidal nanosheets results from the balance between the attraction (van der Waals) and repulsion (electrostatic double layer) potentials the nanosheets are subjected to. If the electrostatic repulsion dominates over the van der Waals attraction, the resulting energy barrier prevents flake aggregation and colloidal stability is observed. In water and organic solvents, the attractive contribution increases with the thickness and is proportional to the surface area of the flakes; as for the electrostatic term, it is mainly thickness-independent but increases with the dielectric constant and is proportional to the surface area of the nanosheets too. As a consequence, for a stable colloid made of a given solvent and flakes of a given thickness distribution, the energy barrier against aggregation is proportional to the surface area of the nanosheets, which implies that large nanosheets are better stabilized than small ones. That seems to be the case in our samples. In the polar solvent mixture, due to a high dielectric constant, the repulsive interaction is so strong that no difference in colloidal stability is observed between the samples. Additional stabilizing solvation effects can also exist in the polar medium due to its particular coordination abilities and possible ion-dipole interactions between the charged nanosheets and the solvent. In the non-polar solvent, the dramatic decrease of the repulsive term (low $\varepsilon_r$) with respect to van der Waals attraction makes the energy barrier far smaller, but still proportional to nanosheet lateral dimensions (if we neglect the effect of the thickness). As a consequence, the flakes are more likely to aggregate, and most of all the smallest ones. Finally, the above mentioned (see the 'phase transfer' section) dispersive interactions probably existing between $MoS_2$ and $CHCl_3$, which can be assimilated to solvation effects, would also increase with nanosheets lateral dimensions and further contribute to the observed better stability of the largest nanosheets.

**Stability of $MoS_2$ nanosheet suspensions in chloroform**

As the phase transfer we present is principally aimed at making easier $MoS_2$ nanosheet functionalization or processing in hydrophobic solvents, characterizing and understanding the corresponding colloidal stability is a particularly relevant issue. For that purpose, we studied the sedimentation process of sample '30 min – B' in chloroform as representative example, and monitored the evolution of the colloid extinction properties as a function of time. The results are presented in Figure 5.

For the first 40 min, the extinction spectrum remains almost identical, indicating the colloid is perfectly stable within such a range of time (Figure 5a). The overall extinction then starts to decrease, quite rapidly in the course of the first hours, and more slowly after 6-7 hours (Figure 5b). In the meanwhile, the corresponding colloidal suspension gets clearer and clearer, along with the formation of a black sediment. The sedimentation process was characterized quantitatively using as above the extinction at 350 nm as a measure of the concentration in suspended $MoS_2$ (Figure 5c). The data set is well fit by an exponential decay curve, which confirms the sedimentation is correctly approximated by a first-order process. This lets us determine a half-life time of 7.2 h for the $MoS_2$-chloroform colloid (time after which half part of the initially suspended material remains in suspension). Such a value is relatively low if compared to other half-life times estimated in polar solvents.[26] Nevertheless, remembering chloroform is a weakly dispersing and coordinating medium for $MoS_2$ nanosheets, this result is quite remarkable. Most of all, it makes $MoS_2$-chloroform colloid appropriate to perform functionalization in hydrophobic organic media or to fabricate devices based on liquid-phase exfoliated nanosheets.

Interestingly, when sonicating back the sample to try and redisperse $MoS_2$ nanosheets after the sedimentation experiment, we noticed the suspension recovers its spectral properties (Figure 5b, d), making it reusable. Note that, due to partial chloroform evaporation, the absolute extinction spectrum of the redispersed colloid (Figure 5b) indicates a slight increase in the sample concentration; nevertheless, the volume variation representing only a 4% decrease over 11 h, it is negligible in our evaluation of $MoS_2$ concentration based on direct extinction measurements over time.

Finally, the normalization at 350 nm of the different extinction spectra measured let appear a typical feature as time goes by: an increase of the scattering contribution (evident beyond 700 nm) along with a red-shift of *A*, *B* and *C* excitonic bands. As discussed in the preceding section, this indicates the flakes ultimately remaining in suspension are the largest one, which is another proof that chloroform stabilizes better large nanosheets than small ones.

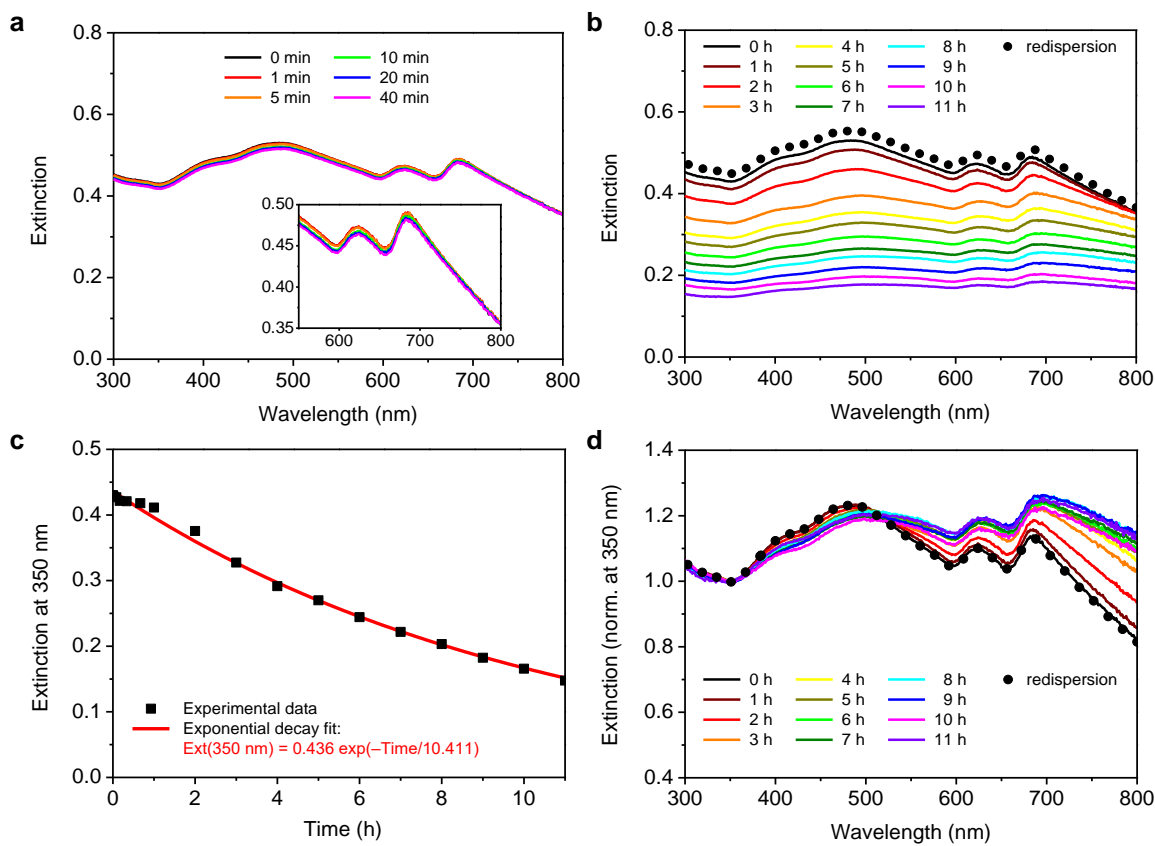

**Figure 5. Stability of MoS$_2$ nanosheet colloidal suspension in chloroform. a.** Extinction spectra of sample '30 min – B' in colloidal suspension in chloroform as a function of time, from 0 min to 40 min (inset: zoom in the 550 nm – 800 nm wavelength domain). **b.** Extinction spectra of sample '30 min – B' in colloidal suspension in chloroform as a function of time, from 0 h to 11 h, and after redispersion by 30 s sonication. **c.** Evolution of the extinction of sample '30 min – B' at 350 nm as a function of time. **d.** Spectra shown in **b** normalized at 350 nm.

**Fluorescence**

As mentioned in the introduction, the ultimate exfoliation of MoS$_2$ leads to monolayer nanosheets that exhibit fluorescence, despite a very low quantum yield, typically below 1%.[2, 47] To determine the influence of the phase transfer on this property we characterized both sets of samples (iPrOH/H$_2$O 7/3 and CHCl$_3$ suspensions) by fluorescence spectroscopy, detecting the typical emission of MoS$_2$ flakes at 650 nm (Figure 5). Note that in our measurement conditions and with liquid-phase exfoliated samples we did not detect any solvatochromic effect[48] on the photoluminescence. Other studies on MoS$_2$ in the liquid phase resulted in the same observation.[14]

In the polar series (Figure 6a-b), no significant change can be noted in the band relative intensities, that is, in the respective single-layer amount between the different samples. Since the probe-type exfoliation has not been optimized toward a specific enrichment in the monolayer content, this is not surprising. Additionally, the very weak emission signal detected from MoS$_2$ fluorescent species does not permit to identify differences that would not be substantial. The same observation is to be done within the non-polar series (Figure 6c-d): the variations in fluorescence intensity as a function of the sample are limited to measurement uncertainty.

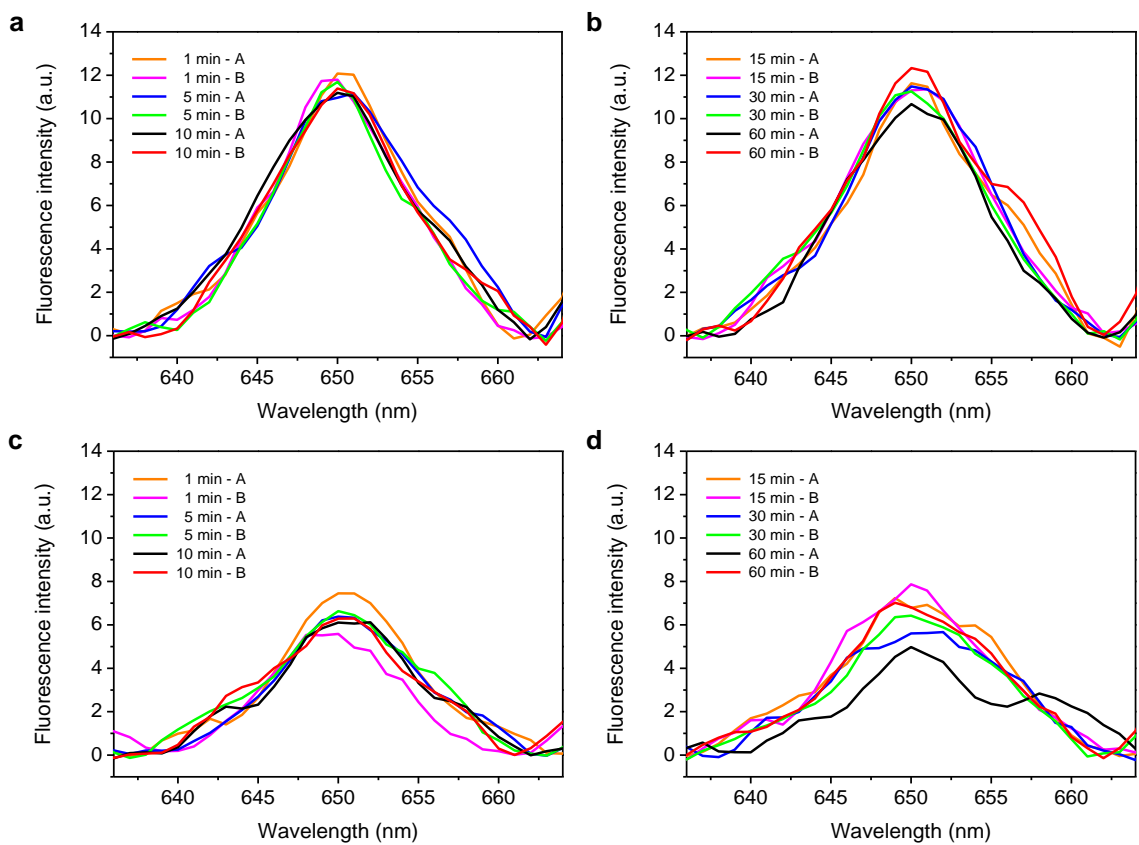

**Figure 6. Fluorescence spectroscopy of exfoliated MoS$_2$ nanosheets.** Fluorescence spectra of obtained colloidal MoS$_2$ nanosheets in suspension in *i*PrOH/H$_2$O 7/3 after the following sonication times: **a.** 1 min (orange and pink lines), 5 min (blue and green lines) and 10 min (black and red lines). **b.** 15 min (orange and pink lines), 30 min (blue and green lines) and 60 min (black and red lines). **c.** and **d.** Fluorescence spectra of obtained colloidal MoS$_2$ nanosheets in suspension in chloroform after the phase transfer. As in **a** and **b**, each sample is referred to as the sonication time used for the initial exfoliation experiment in *i*PrOH/H$_2$O 7/3, associated with the letter A or B, indicating the first and second replicate respectively. All spectra are baseline-corrected.

Of more interest is the comparison of the polar and non-polar suspensions. In similar excitation and measurement conditions, chloroform-MoS$_2$ colloids exhibit approximately half of the fluorescence of the initial *i*PrOH/H$_2$O dispersions. Such a decrease in the photoluminescence intensity indicates the single layer content drops after the phase transfer. This confirms the preceding conclusions drawn from the extinction observations, namely the nanosheet partial aggregation occurring in chloroform.

**Conclusions**

The work reported here shows that colloidal MoS$_2$ nanosheets prepared in a polar medium can be transferred to a non-polar solvent such as chloroform, provided a careful and stepwise removal of the polar solvent(s) is performed. Additionally, within the six different nanosheet batches tested, the phase transfer is operative whatever the distribution size and proved to be reproducible. UV-Visible analysis evidenced partial aggregation of the flakes in chloroform (confirmed by fluorescence measurements) that is mainly controlled by their lateral size, the largest nanosheets being the most stable. Nevertheless, the corresponding mid-term colloidal stability, with a typical half-time of *ca.* 7 h and easily retrievable, is perfectly suited for functionalization experiments of MoS$_2$ nanosheets with hydrophobic molecules or for their manipulation within low-boiling-point non-polar solvents; especially as no possibly competing or contaminating surfactant was used to help flake stabilization in suspension.

**Experimental Section**

**General**

MoS$_2$ powder (< 2 μm, 99%) was purchased from Sigma Aldrich. Solvents were purchased from Scharlau chemicals and used as received; water was obtained from a Milli-Q filtration station ("Type 1" ultrapure water; resistivity: 18.2 MΩ.cm at 25 °C).

**Liquid-phase exfoliation**

MoS$_2$ powder (< 2 μm, 99%, 1 mg.mL$^{-1}$) was dispersed in 80 mL of a 7/3 mixture (v/v) of isopropanol (*i*PrOH) and water, in a 100 mL round-bottom flask further cooled down using an ice/water bath. The liquid-phase exfoliation was performed using an ultrasonic probe (Vibracell$^{TM}$ 75115, Bioblock Scientific, 500 W) immersed in the dispersion and operating at the amplitude of 40%. After sonication, the black suspension was distributed into six 20 mL glass vials that were centrifuged for 30 min at 990 *g* (3,000 rpm, Allegra$^®$ X-15R Beckman Coulter centrifuge, FX6100 rotor, 25 °C). The corresponding olive-color supernatants (~10 mL) were carefully separated from the black sediment and collected each in another 20 mL vial for further phase transfer.

Six different experiments were carried out, differing in the sonication time (continuous ultrasound irradiation): 1 min, 5 min, 10 min, 15 min, 30 min and 60 min, respectively. Each experiment was replicated.

**Phase transfer of the nanosheets**

In a typical experiment, chloroform (5 mL) and water (5 mL) were added to the previously collected colloidal suspension (10 mL) and the mixture was shaken a few seconds. Solvent demixing caused the nanosheets to flocculate at the interface between the polar and non-polar phases. Both phases were removed and the remaining nanosheets were redispersed in pure *i*PrOH (2 mL). This procedure was repeated over the six vials corresponding to a same experiment and precipitated nanosheets were accumulated in the above-mentioned 2-mL *i*PrOH redispersion. Chloroform was added (1 mL) and the nanosheets were separated from the solvent via centrifugation (Hettich Mikro 120 centrifuge, 24-tube rotor, 14,000 rpm, 18,626 *g*, 10 min). The resulting black pellet was washed another time with *i*PrOH/CHCl$_3$ 2/1 (v/v, 1 mL), *i.e.* redispersed in the solvent mixture by a few-second bath sonication (Fisherbrand FB15051 bath sonicator, ultrasound frequency 37 kHz, 280 W, ultrasonic peak max. 320 W, standard sine-wave modulation), then centrifuged (Hettich Mikro 120 centrifuge,

18,626 *g*, 10 min). Finally, the black sediment was washed four times with pure chloroform (sonication/centrifugation sequences using the bath sonicator a few seconds and the Hettich Mikro 120 centrifuge at 16,060 *g* for 7 min) and redispersed in chloroform.

**UV-Visible spectroscopy**

The extinction spectra were measured in a quartz cuvette (path length = 1 cm) with a Cary 50 UV-Visible spectrophotometer. In the case of the chloroform suspensions, the spectra were recorded immediately after a few-second sonication to ensure complete redispersion of the nanosheets.

**Sedimentation/redispersion experiment**

A 3-mL suspension of $MoS_2$ nanosheets in chloroform (second replicate prepared by 30 min sonication in *i*PrOH/$H_2O$ 7/3, v/v) was sonicated for 2 min and transferred into a quartz cuvette (path length = 1 cm). The cuvette was closed with a stopper and the extinction spectrum of the suspension was measured immediately (time zero measurement). The suspension was left at room temperature and its extinction spectrum was recorded at various time intervals. After 11 h monitoring, the sample was stirred and sonicated for 1 min. Its extinction spectrum was recorded and compared to the time zero measurement.

**Transmission Electron Microscopy**

Colloidal samples '1 min – A' and '60 min – A', as-prepared and after phase transfer, were respectively drop-casted onto 200 square mesh copper grids covered with a carbon film. They were observed using a JEOL JEM 2100 microscope operated at 200 kV.

**Raman spectroscopy**

Colloidal samples '1 min – A' and '60 min – A', as-prepared and after phase transfer, were respectively drop-casted and dried onto glass slides at 50 °C. Their Raman spectra were recorded with a Bruker Senterra confocal Raman microscope (Bruker Optic, Ettlingen, Germany, resolution 3-5 $cm^{-1}$) using the following parameters: objective NA 0.75, 50×; laser excitation: 532 nm, 2 mW. Each spectrum results from the average of 10 measurements carried out in different regions distributed all over the sample.

**Fluorescence spectroscopy**

Photoluminescence (PL) spectra were performed on a Fluorolog®-3 HORIBA spectrofluorometer. All PL spectra were recorded at an excitation wavelength $\lambda_{exc}$ = 412 nm (integration time: 1 s; excitation and detection slits: 5 nm bandpass; grating: 1200 blaze 500). The samples were diluted so that they exhibit the same extinction at the excitation wavelength and an extinction below 0.1 at the emission wavelength, in order to avoid possible re-absorption.

**Acknowledgements**

E.G. gratefully acknowledges the AMAROUT II fellowship program for receiving a grant for transnational mobility (Marie Curie Action, FP7-PEOPLE-2011-COFUND (291803)). E.M.P. acknowledges financial support from the European Research Council (MINT, ERC-StG-307609) and from the MINECO of Spain (CTQ2014-60541-P), as well as from the European Union structural funds and the Comunidad de Madrid through MAD2D-CM program (S2013/MIT-3007).

**Keywords:** molybdenum disulfide • liquid-phase exfoliation • ultrasonic probe • phase transfer • colloidal stability

**Table of Contents**

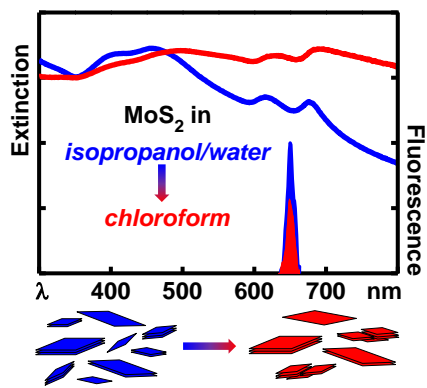

*Emerson Giovanelli,* Andres Castellanos-Gomez and Emilio M. Pérez**

*Page No. – Page No.*

**Surfactant-Free Polar-to-Non-Polar-Phase Transfer of Exfoliated MoS$_2$ Two-Dimensional Colloids**

# Supporting Information

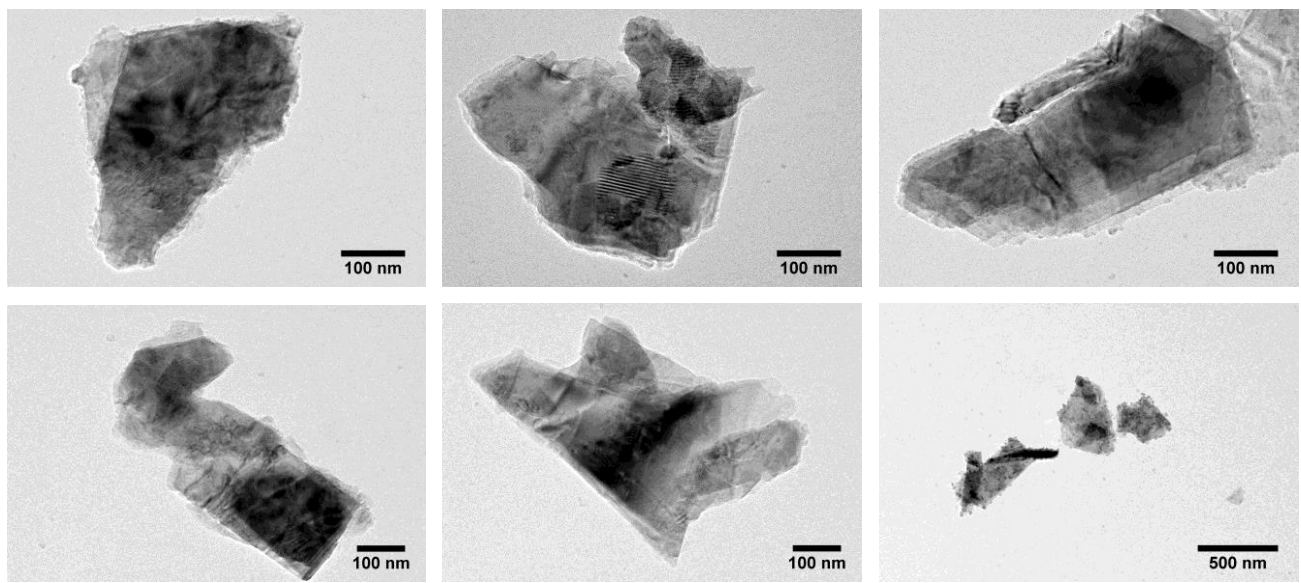

**Figure S1.** TEM micrographs of as-obtained nanosheets from sample '1 min – A' after the exfoliation-centrifugation process in *i*PrOH/H$_2$O 7/3.

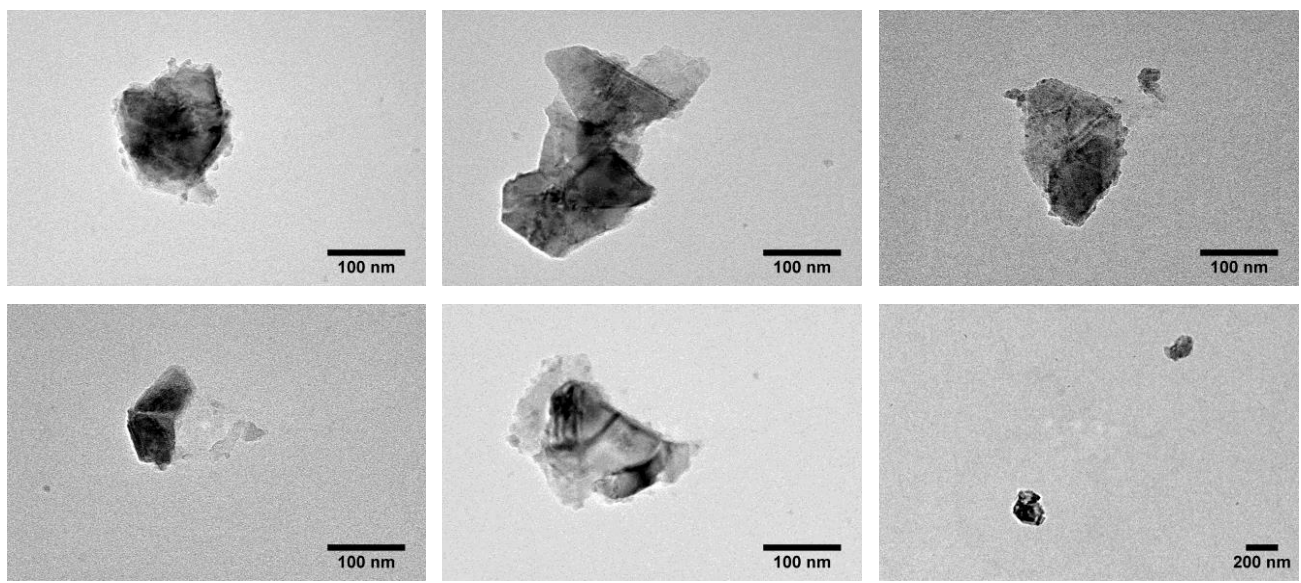

**Figure S2.** TEM micrographs of as-obtained nanosheets from sample '60 min – A' after the exfoliation-centrifugation process in *i*PrOH/H$_2$O 7/3.

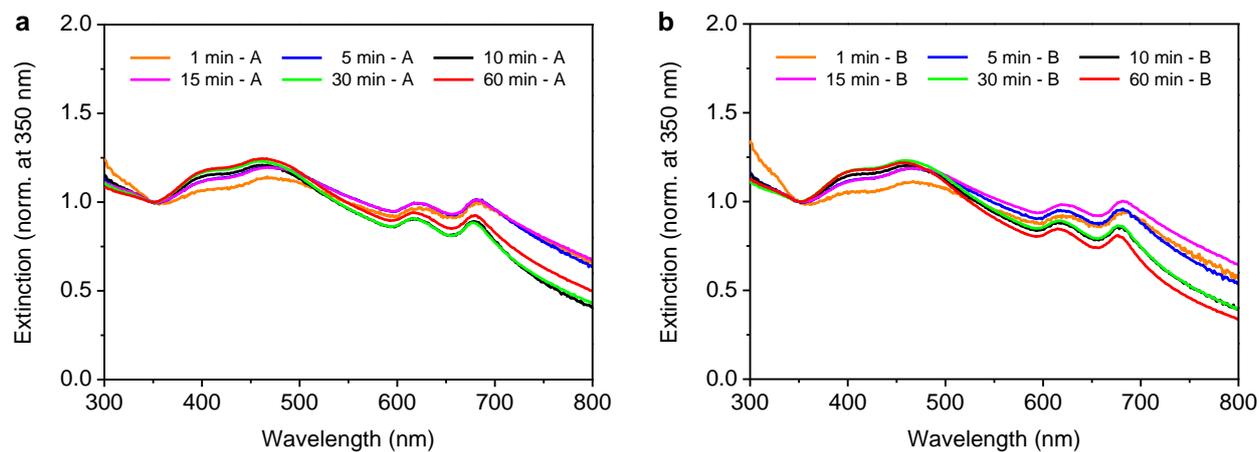

**Figure S3. UV-Visible spectroscopy of MoS$_2$ colloids obtained after exfoliation of MoS$_2$ powder dispersions in iPrOH/H$_2$O 7/3 (v/v) by probe sonication, spectra mormalized at 350 nm. a.** Series of replicates A. **b.** Series of replicates B. Each sample is referred to as the sonication time used for the exfoliation experiment in *i*PrOH/H$_2$O 7/3, associated with the letter A or B, indicating the first and second replicates respectively.

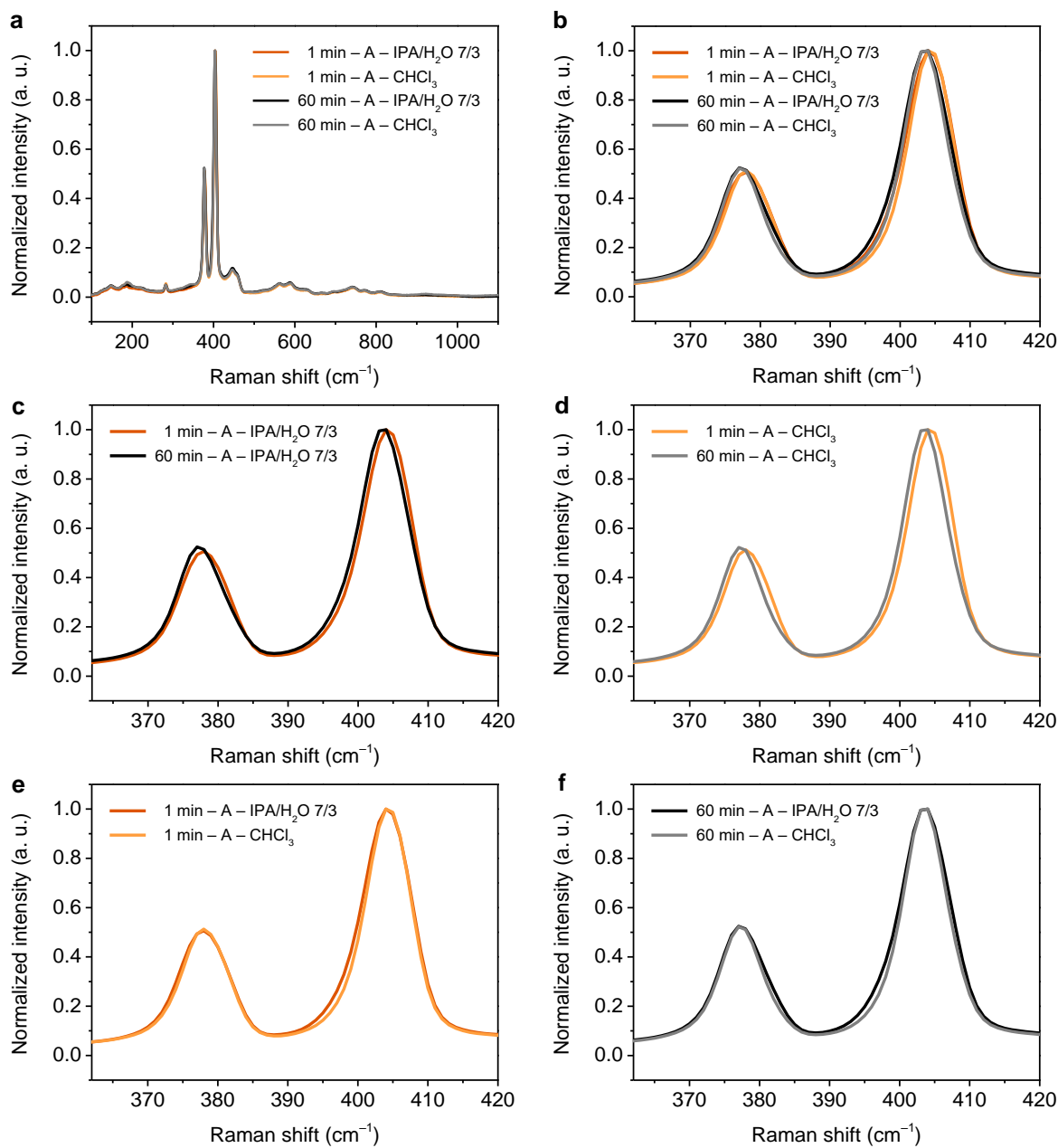

**Figure S4. Raman spectroscopy of MoS$_2$ nanosheets obtained after exfoliation of MoS$_2$ powder dispersions in *i*PrOH/H$_2$O 7/3 (v/v) and after transfer into chloroform.** Each sample is referred to as the corresponding sonication time, the letter A indicating the measurements were carried out on the first replicates of each experiment. **a.** Raman spectra of as-obtained nanosheets in *i*PrOH/H$_2$O 7/3 after 1 min probe sonication (dark orange), after 60 min probe sonication (black), and corresponding nanosheets after the phase transfer to chloroform (light orange and grey, respectively). **b-f.** Zoom of the spectra shown in **a.** in the 362-420 cm$^{-1}$ region, showing MoS$_2$ main Raman bands: **b.** All spectra; **c.** Spectra of the nanosheets obtained in *i*PrOH/H$_2$O 7/3 after 1 min and 60 min probe sonication; **d.** Spectra of the nanosheets obtained in CHCl$_3$ after phase transfer of the preceding samples; **e.** Spectra of the nanosheets obtained from a 1 min probe sonication, before and after phase transfer; **f.** Spectra of the nanosheets obtained from a 60 min probe sonication, before and after phase transfer.

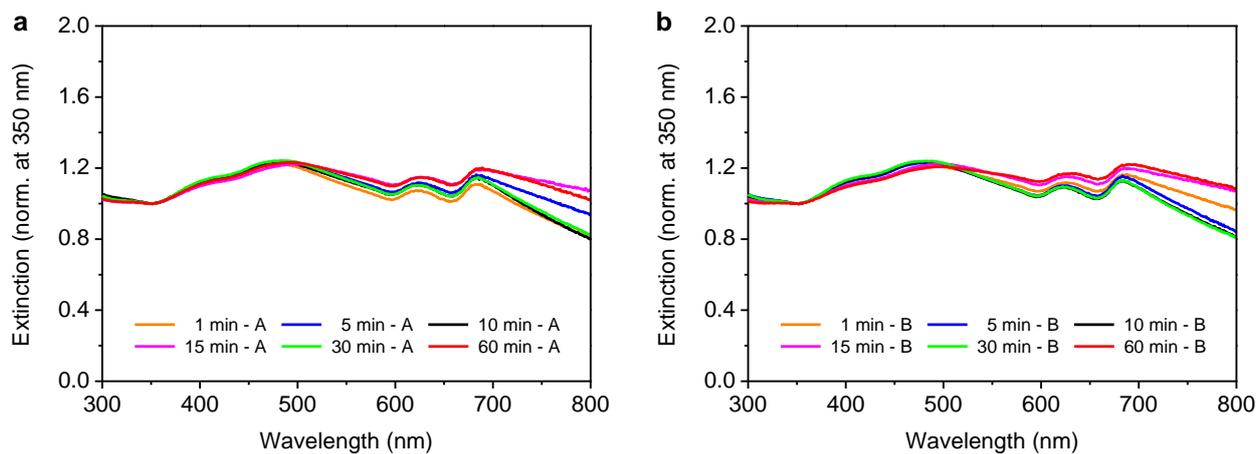

**Figure S5.** UV-Visible spectroscopy of exfoliated MoS$_2$ nanosheets after phase transfer into chloroform, spectra mormalized at 350 nm. **a.** Series of replicates A. **b.** Series of replicates B. Each sample is referred to as the sonication time used for the initial exfoliation experiment in iPrOH/H$_2$O 7/3, associated with the letter A or B, indicating the first and second replicate respectively.

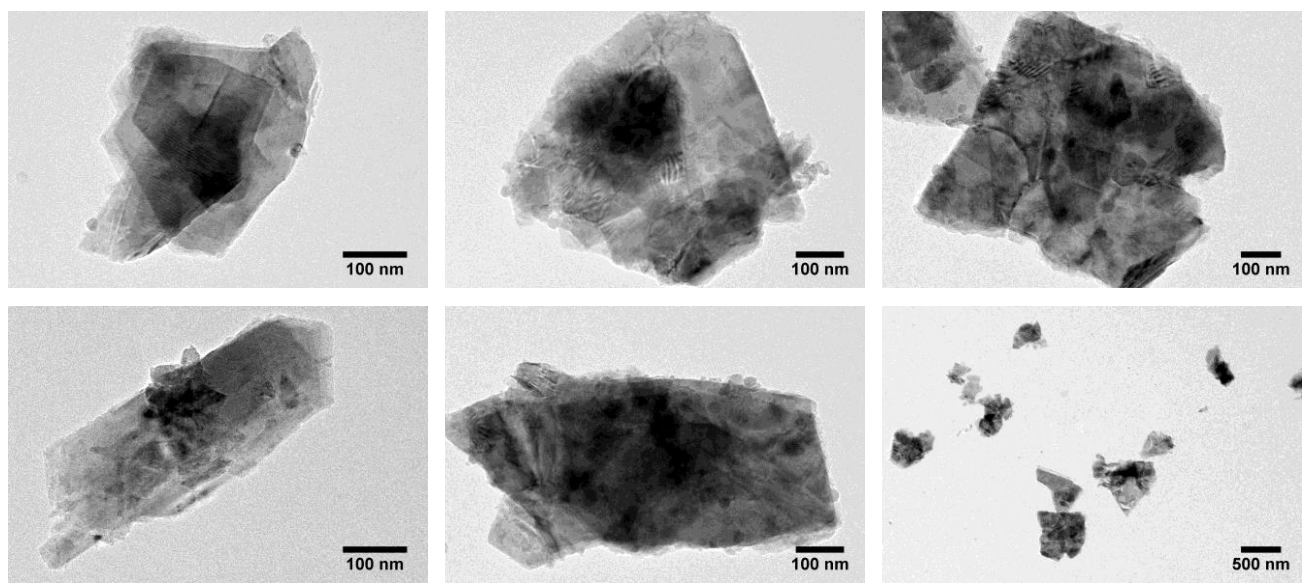

**Figure S6.** TEM micrographs of nanosheets from sample '1 min – A' after the phase transfer into CHCl$_3$.

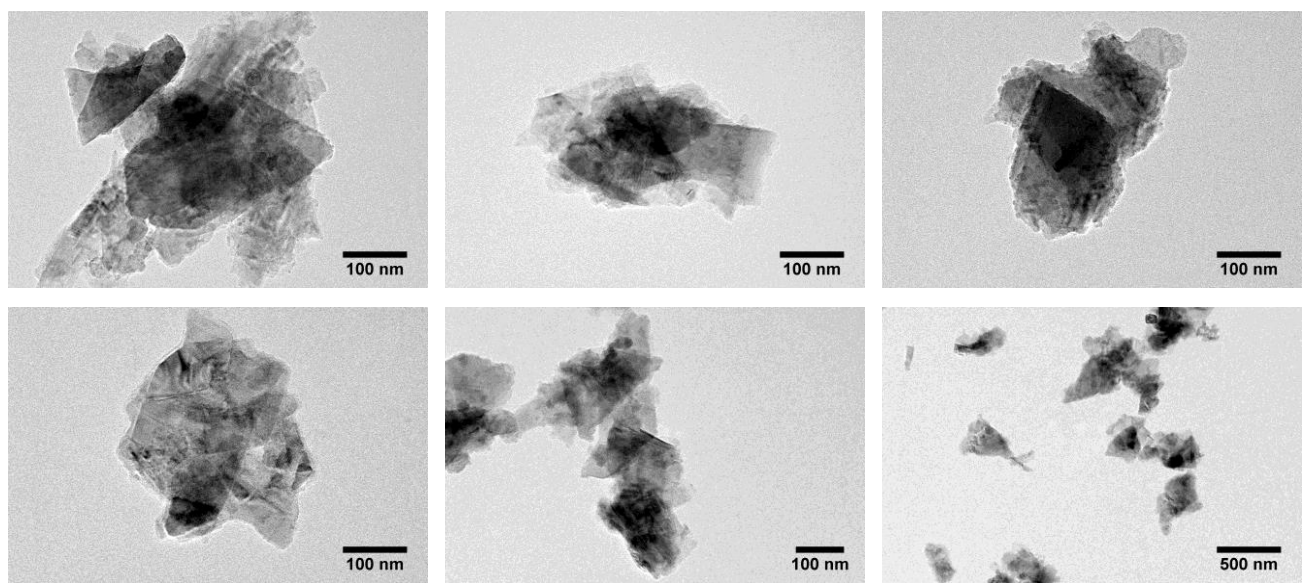

**Figure S7.** TEM micrographs of nanosheets from sample '60 min – A' after the phase transfer into CHCl$_3$.